# A Critical Evaluation of a Self-Driving Laboratory for the Optimization of Electrodeposited Earth-Abundant Mixed-Metal Oxide Catalysts for the Oxygen Evolution Reaction (OER)

Erfan Fatehi[1], Manish Thadani[1], Gabriel Birsan[2], Robert W Black[1*]

1. National Research Council of Canada, Energy, Mining, and Environment Research Centre, 2620 Speakman Drive, Mississauga, ON, L5K 1B4, Canada

2. Natural Resources Canada, Canmet MATERIALS, Innovation and Energy Technology Sector, 183 Longwood Road South, Hamilton, Ontario, L8P 0A5, Canada

*Robert.black@nrc-cnrc.gc.ca

*Abstract*

This work highlights the potential of earth-abundant mixed-metal oxide catalysts for the acid-based oxygen evolution reaction. These catalysts offer numerous combinations of metal-centre compositions, which can enhance catalytic activity and stability compared to precious-metal-based catalysts commonly used today. Despite substantial research in this field, there is a need for new methods and approaches to accelerate the exploration of these materials. In this study, we present a comprehensive approach to designing, developing, and implementing a self-driving laboratory to optimize the electrodeposition synthesis of amorphous mixed-metal oxide catalysts for the acidic oxygen evolution reaction. We particularly emphasize the development of methodologies to address experimental variability. We investigate crucial parameters and considerations when transitioning from manual bench-top synthesis and evaluation to automation and machine learning guided optimization. We address both experimental and optimization algorithm considerations in the presence of experimental variability. To illustrate our approach, we demonstrate the optimization of $CoFeMnPbO_x$ electrodeposited catalyst materials through multiple campaigns. Our results highlight considerations for optimizing overpotential and stability based on the outcomes of our experiments.

*Introduction*

As the world faces the pressing challenge of climate change, the development of clean and sustainable energy sources is more important than ever. One promising avenue is the use of renewable energy technologies, such as water splitting, to generate hydrogen as a fuel. This process requires a highly efficient and stable catalyst for the oxygen evolution reaction (OER), which is often expensive and relies on precious metals such as platinum or iridium for acceptable performance in acidic OER environments in terms of catalytic activity and stability.[1,2] Alternatives to these catalysts are not yet available that match these catalysts' performance for activity and acidic stability.[3,4] Research on non-noble metal mixed oxide catalysts, in particular those based on earth-abundant 3d transition metals for acidic OER, has gained significant attention as a potential solution to this challenge.[5,6] However, in acidic media these transition metal oxides are quite vulnerable under oxidative conditions, demonstrating poor activity and stability compared to catalysts based on precious metals.

Extensive research has been conducted to assess and discover catalyst materials with promising properties. However, there is still a vast array of potential candidates in this field that we have only begun to comprehend and exploit. The sheer number of material combinations, parameters, and synthesis conditions adds to the extensive data space, necessitating thoughtful deliberation to expedite the production and evaluation of these materials. To perform this optimization, we developed and deployed a self-driving laboratory, in collaboration with North Robotics.[7] Self-driving laboratories represent a new paradigm in how research is performed, combining laboratory automation with machine learning optimization to accelerate the pace at which experimentation is performed, often leading to 10-100 times increase in the throughput and efficiency towards optimization of a new material or process.[8–10] To date, numerous studies have explored the concept of high-throughput experimentation to explore the vast and rich space of mixed-metal oxide catalysts for OER across all media (acidic, neutral, and alkaline).[11–16] High-throughput experimentation looks to parallelize the synthesis and characterization process to produce data that is reliable, accurate, and in abundance to down-select optimal materials based on the search spaces. The result of such an exploration is a rich dataset of material activity, which can then be used to feed various other sources of information, further driving the material discovery process. This approach has also been used outside of OER catalysis and is a powerful tool to explore a variety of electrocatalysis processes such as ORR[17] and $CO_2R$.[18,19]

Self-driving laboratories offer an alternative approach to high-throughput experimentation, utilizing probabilistic optimization algorithms, such as Bayesian optimization and genetic algorithms [20], to efficiently optimize the experimental process in a limited number of experiments and/or a potentially more efficient exploration of the design space through the use of machine learning guided optimization. This becomes critically important when dealing with high dimensional and/or large amounts of explorable parameters, as the design space can quickly outgrow the feasible exploration space possible using high throughput experimentation alone. The first example of a self-driving laboratory for material applications was deployed for the optimization of carbon nanotube growth.[21] Since then, self-driving laboratories have grown in popularity to address a plethora of various material domains, including but not limited to organic molecule synthesis,[22,23] structured materials,[24] photovoltaics,[25] thin-film materials,[26–28] nanomaterials,[29,30] and electrolytes.[31,32] These works have demonstrated the success of self-driving laboratories to optimize a synthesis, characterization, or process, while also demonstrating flexibility in the application of this concept towards handling a variety of different scientific challenges.

This study focuses on amorphous transition mixed metal oxides and optimizes a set of experimental conditions to identify optimal regions of potential $CoFeMnPbO_x$ catalysts as it relates to both activity (minimize overpotential) and stability utilizing a self-driving laboratory. This study serves as a guide towards developing a self-driving laboratory for electrocatalysis applications and presents our critical evaluation, findings, considerations, and suggestions when looking to deploy a self-driving laboratory for electrodeposition and electrocatalysis evaluation. In particular, we explore various approaches to the design of the self-driving laboratory, the importance of reproducibility and validity in the data, proxy experimentation, experimental noise and variance on the metric of optimization, and finally culminate this into the optimization of a mixed-metal oxide catalyst for the OER reaction.

We set out to design a self-driving laboratory for the electrochemical synthesis and characterization of the mixed-metal oxide catalyst material, with the goal of a) optimizing the synthesis conditions of the mixed-metal oxide electrodeposition to achieve a material with the lowest OER overpotential within the defined experimental space, and b) elucidate stability information from these catalysts through the utilization of proxy experimentation. We focus our efforts on OER active earth-abundant elements (Co, Fe) coupled with a structural element (Mn, Pb) synthesized via electrodeposition. Electrodeposition allows for the fabrication of uniform thin films with good adhesion and electrical contact with the substrate (in this case, fluorinated tin oxide FTO), along with the ability to directly probe the catalyst activity without influence from any geometric and/or size distribution influences that may arise when working with non-thin film morphologies (eg. power materials). For this particular challenge, previous research has demonstrated the effectiveness of anodic electrodeposition to combine catalytically active transition metal species, such as Co, Fe, and Ni, with those of more acidic stable, albeit low-activity species such as Pb or Bi.[33,34] In previous work, Hyunh *et. al.* approached templating active 3d transition elements such as Co and Fe with the acid-stable oxide $PbO_x$ to produce $CoFePbO_x$ via anodic electrodeposition in near-neutral deposition conditions.[35] This work was further extended to look at the stability of templated $NiPbO_x$ and $NiFePbO_x$, demonstrating success in stabilizing the $NiO_x$-based compound for extended periods (>20 h) under acidic OER conditions.[36] $NiO_x$ traditionally exhibits extremely poor stability under acidic OER conditions, but the inclusion of Fe and Pb into the $NiO_x$ network can modulate the local acidity and the role of an acid-stable supporting matrix to extend the catalyst centre lifetime (ie. reduce the rate of dissolution). In another work, $Pb^{2+}$ ions were electrodeposited with $Co^{2+}$ and $Fe^{3+}$ ions under acidic conditions and demonstrated the intrinsically stable mixed-metal oxide through an oxidation–redeposition equilibrium formed *in-situ* during OER.[37] Both of these examples demonstrate the concept of 'self-healing' catalysts, a concept introduced by Nocera,[38,39] to describe the nature of these catalysts and their inherent activity and stability under acidic water splitting conditions based on a dynamic reconstruction/redeposition process. In this work, we were heavily influenced by the experimental protocol of Chatti *et. al.* to electrodeposit in acidic media.[37] They demonstrated that the mixed-metal oxide could be formed in acidic conditions through an applied anodic potential, but only in the presence of sufficient structural elements or risk the immediate dissolution of the active material (eg. $CoO_x$ vs. $CoPbO_x$). Their experimental procedure was an inspiration for this work, and hence we used a modified procedure as the baseline electrodeposition protocol to develop our self-driving laboratory. It is expected that there will be an optimal trade-off between the composition of structural elements vs. active elements – as more structural elements are utilized, the overall catalytic activity (and hence overpotential) will reduce due to the inherent poor OER catalytic activity of PbO$_x$ or MnO$_x$ alone.

*Results & Discussion*

*Framework for Self-Driving Laboratory*

An outline of the workflow for the designed self-driving laboratory is shown in **Figure 1.** The automated platform was designed to i) handle and deposit known concentrations of metal salts dissolved in aqueous media, ii) handle conductive glass slides as the substrate, iii) load and unload these substrates into 3-electrode electrochemical cells for the electrodeposition and electrochemical performance testing. The platform is comprised of two physical modules – solution handling automation, and electrochemistry handling automation. The third module is the data handling and optimization framework to close the experimental loop.

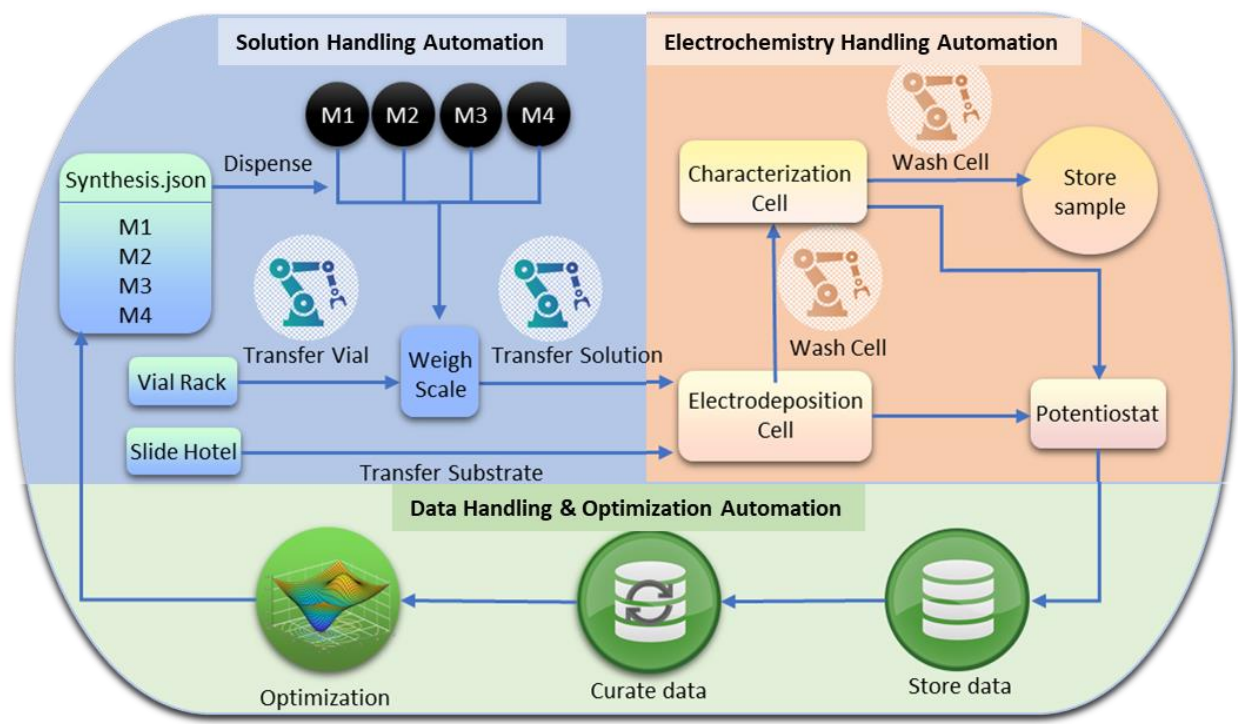

**Figure 1 –** *Schematic of the self-driving laboratory for the electrochemical deposition and performance characterization of amorphous mixed-metal oxide OER catalysts.*

The physical platform is shown in **Figure 2a,** with critical components labelled. The substrate handling is performed with a vacuum gripper attached to a SCARA robot, with the gripper custom designed to not touch the surface of the pre-cleaned substrate materials. The substrates are pre-cleaned and loaded onto a 36-port slide hotel. The solution handling is performed via integrated Cavro XCalibur syringe pumps, a pipette tip attached to the SCARA robot arm, and the SCARA robot arms gripper which can grab, cap, and uncap 4 mL vials. Connected to the syringe pumps are stock solutions of 1M metal nitrate salts in 18 MΩ Milli-Q water. To prepare the electrodeposition solution, a clean vial is loaded onto a weighing scale. Volumes of between 0 – 1 mL of the metal-salt stock solutions are dispensed via 1/16$^{th}$ tubing into the vial, totalling a volume of 1 mL. We can accurately dispense to a precision of 0.02 mL, determined via repeated dispense procedures. Each addition of metal-salt solution is quantified via the change in total volume mass, which is used to determine the final salt concentrations in the solution.

Both the electrodeposition and electrochemical performance evaluation occur in our custom-made automated 3-electrode electrochemical cells (***Figure 2b).*** Two separate cells are used, one for electrodeposition which contains the robot-prepared diluted salt solution (between 0-1 M nitrate solutions in 0.1 M HNO$_3$, see *Supporting Information* for details), and one for performance evaluation in 0.01M H$_2$SO$_4$ electrolyte (pH measured 1.7-1.8 between batches). Two separate cells were utilized to ensure no cross-contamination between electrodeposition and performance characterization protocols. After each deposition or performance evaluation, the cells are washed using successive fill-empty cycles of deionized water. To determine the number of washing cycles that are required to remove trace impurities down to the single ppm level, the cell was filled with 0.01 M H$_2$SO$_4$, and the pH was monitored after each successive wash cycle. As shown in ***Figure S1***, the pH of our test solution tests neutral after ~8 successive washing stages, and thus 10 steps were determined to be an adequate number of washing steps to bring metal contaminants down below the single ppm level to avoid cross-contamination after successive experiments.

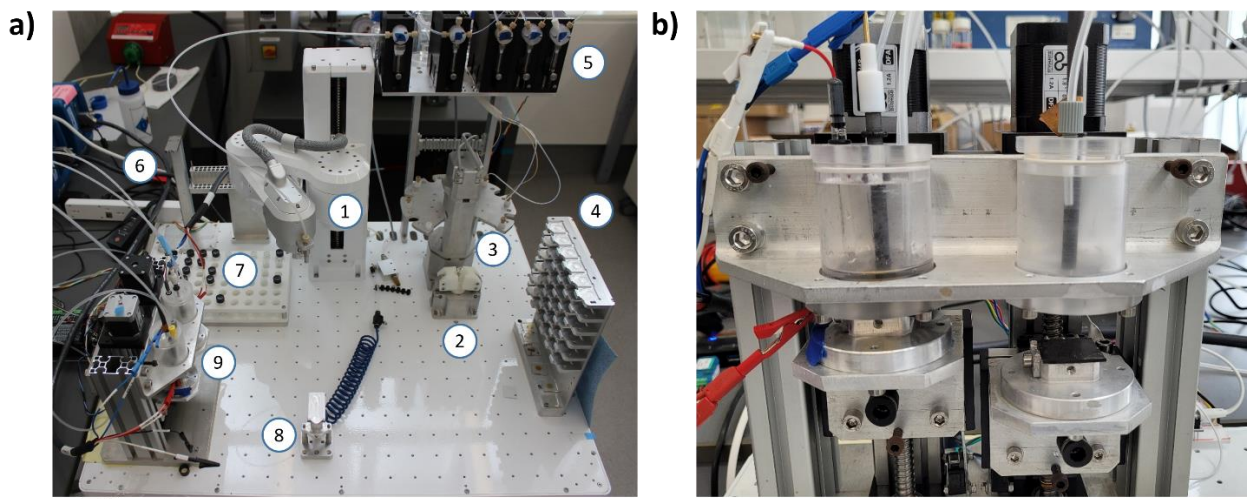

**Figure 2 -** *a) EMAP (Electrocatalysis MAP) comprised of 1) robotic arm with integrated pipette holder, 2) weigh scale and gripper, 3) liquid dispensing carousal, 4) slide hotel, 5) syringe pumps for liquid handling, 6) pipette tip holder, 7) vial rack, 8) slide gripper and 9) automated electrochemical cell. Not shown are the stock solution bottles, which are positioned at the back of the platform. b) Automated electrochemical cell.*

*Electrochemical Performance Characterization Protocol Development*

Critical to this work is the workflow development to properly evaluate the catalyst performance. In particular, our goal is to develop an automated experimental protocol that can be used to optimize the catalyst overpotential, while also gathering information on the material composition relationship to acid stability during OER. As will be discussed in subsequent sections, experimental variability, in particular those related to the desired metric to be optimized, must be understood and sources of error and variance in the experimental protocol must be minimized for a successful optimization process. Furthermore, it is desirable to minimize the total reaction time and increase the throughput of experiments, without losing experimental data fidelity. Typically, to evaluate the performance of an electrocatalyst, techniques such as linear sweep voltammetry (LSV) are used to obtain a metric of overpotential versus a set current rate. Historically, 10 mA/cm$^2$ has been a benchmark target for OER performance.[40,41] These measurements can be performed relatively quickly, assuming the catalyst is in an electrochemically steady state with its environment. For our particular system and desire to optimize non-noble mixed-metal oxides for stability in acidic OER, stability is a great concern. Unfortunately, catalyst stability is typically a metric that can take significant time to measure, most times involving the entire lifetime (dissolution) of the catalyst which can take tens to even hundreds of hours. As such, as part of this work, we developed a characterization workflow that can a) capture metrics of stability and catalyst overpotential in a meaningful, reproducible manner, and b) can be deployed within the confines of our automated platform while also providing rapid measurements.

To evaluate electrochemical performance, we performed 30-minute chronopotentiometry with intermittent electrochemical impedance spectroscopy. ***Scheme 1*** summarizes our performance measurement protocol. As part of this development, we relied on the concept of 'proxy' figures of merit. Proxy metrics involve measurements that elucidate specific performance metrics of a material that may lack fundamental considerations and/or be influenced by external factors limiting the use of these metrics outside of the housed experimental campaign, but is acceptable within the confines of an experimental campaign when utilized to compare one sample to another for optimization purposes. With this reasoning, we chose to evaluate overpotential via constant current hold for 30 minutes. Such a measurement allows for the catalyst to reach a 'steady state' within the electrochemical environment, as would be expected of a mixed-metal oxide in an acidic environment undergoing surface reconstruction upon initial electrification.[42,43] Automation complexities associated with more traditional benchmarking methods that better isolate intrinsic catalyst activity, such as RDE with tafel analysis, do prevent such methods from being performed and utilise 'true' catalyst performance as a benchmark of optimization. However, chronopotentiometry is a quantitative metric that can be used to evaluate catalyst performance, and within the confines of our optimization, this will allow the optimizer to make informed decisions regarding relative catalyst performance, and optimize the synthesis conditions as such.[44]

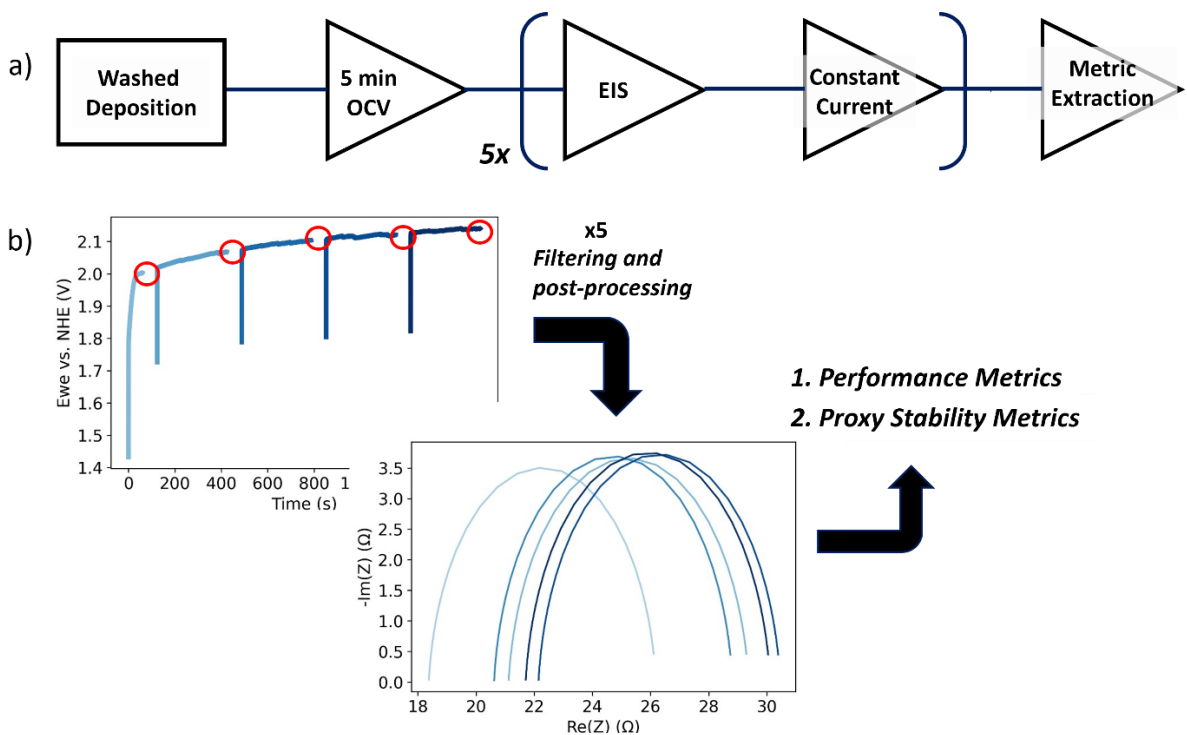

**Scheme 1 -** *a) Scheme for the electrochemical evaluation of the deposited OER catalysts, b) depiction of the resulting electrochemical profile.*

To properly evaluate catalyst stability, the most appropriate measurements involve the direct measurement of the catalyst material dissolution from the material into the electrolyte, via methods such as quartz crystal microbalance or inductively-coupled plasma mass spectrometry.[45–47] These measurements require either expensive infrastructure and/or precise measurements which can be difficult to perform within our self-driving laboratory. A *proxy* method of catalyst stability measurement using EIS coupled with chronopotentiometry was developed to evaluate the validity of a *proxy* predictive model of catalyst lifetime. To test this approach, we generated over 20 electrodeposited Mn/Co/NiO$_x$-based catalyst materials that were reproduced using similar electrodeposition procedures as outlined by Hyun *et. al.*[35] Using a low constant applied current (0.1 mA/cm$^2$) coupled with intermittent EIS measurements, we were able to fit Randles circuit and extract values for R$_{ct}$ and the constant phase element (CPE) (to which was extracted the effective capacitance C$_{eff}$) as a function of time, and correlate these changes to the true lifetime of the catalyst. A more elaborate discussion of this process is provided in the *Supporting Information*. To summarize, we determined that for our samples of interest, there is a near-linear decay in the effective capacitance of the sample versus time under constant current, which can be used to linearly extrapolate and predict the catalyst end of life after operating for only 20 minutes. We understand that a much more comprehensive study is necessary to gather a larger data set to gain more accuracy and test the thresholds of our predictive model, which are the focus of future studies. Furthermore, it is worth exploring and developing other methods of accelerated stress testing of OER catalysts, which can be used as strong predictors of catalyst lifetime when coupled with ML to elucidate performance-prediction models. However, for our purposes in determining an adequate OER performance characterization protocol within the confines of our system (electrodeposited non-noble metal OER thin film catalysts), the presented measurements serve as an adequate *proxy* indicator of catalyst lifetime to compare those catalyst synthesized from the self-driving laboratory.

*Electrodeposition Protocol Development*

A standard electrodeposition protocol is required to establish an acceptable level of sample-to-sample reproducibility. We used a modified version of Chatti *et. al.*[37] electrodeposition procedure as our baseline synthesis method. Modifications to this procedure include using 1M $HNO_3$ as the supporting electrolyte to eliminate any possibility of sulphate-precipitates, in particular $PbSO_4$, which were observed to occur when sulphuric acid was used as the supporting electrodeposition electrolyte. For this particular study, the electrochemical deposition profile remains constant, and the electrodeposition solution environment (precursor cation and concentration) are the dependent variables, with voltage and time remaining constant to simplify the automation complexity at this stage. In future work, we plan to deploy this SDL to optimize the electrodeposition voltage and time as part of the Bayesian optimization protocol. The electrodeposition voltage must be selected to ensure that there is enough electrochemical driving force to electrodeposit the metals of interest in their oxy/hydroxyl state at sufficient rates, while also limiting the oxygen evolution reaction. To determine an adequate electrodeposition voltage, multiple repeat $CoPbO_x$ electrocatalysts were produced with applied deposition potentials of 1.7 V, 1.9 V, 2.1 V, and 2.3 V vs. NHE for 3 minutes (*Figure 3a*), followed by OER performance evaluation (*Figure 3b*). The current response during deposition increases with subsequent electrodeposition voltage increases, as would be expected. Images of the deposited samples after electrodeposition indicates that the degree of bubble templating is greater at more positive deposition voltages, which corresponds to OER. (*Figure S5*). Post-deposition OER performance evaluation in 0.01 M $H_2SO_4$ reveals that a deposition potential of 1.7 V vs. NHE results in poor catalyst activity, evident from a greater overpotential as well as poor reaction kinetics (determined from $R_{ct}$) and effective capacitance (from $C_{eff}$) as analyzed from the EIS data (***Figure 3c*** and ***Figure 3d***, respectively). Deposition potentials of 1.9 V vs. NHE or greater result in an appreciable OER performance. Directly measuring the catalyst OER performance overpotential remains the same at 1.88 V vs. NHE, irrespective of the electrodeposition voltage, and the $R_{ct}$ and $C_{eff}$ show only minor changes. We conclude that for this study, 1.9 V vs. NHE is an acceptable electrodeposition voltage, whereas any greater voltage will have a large propensity for sample-to-sample variability given the increased bubble templating due to the competing OER during electrodeposition.

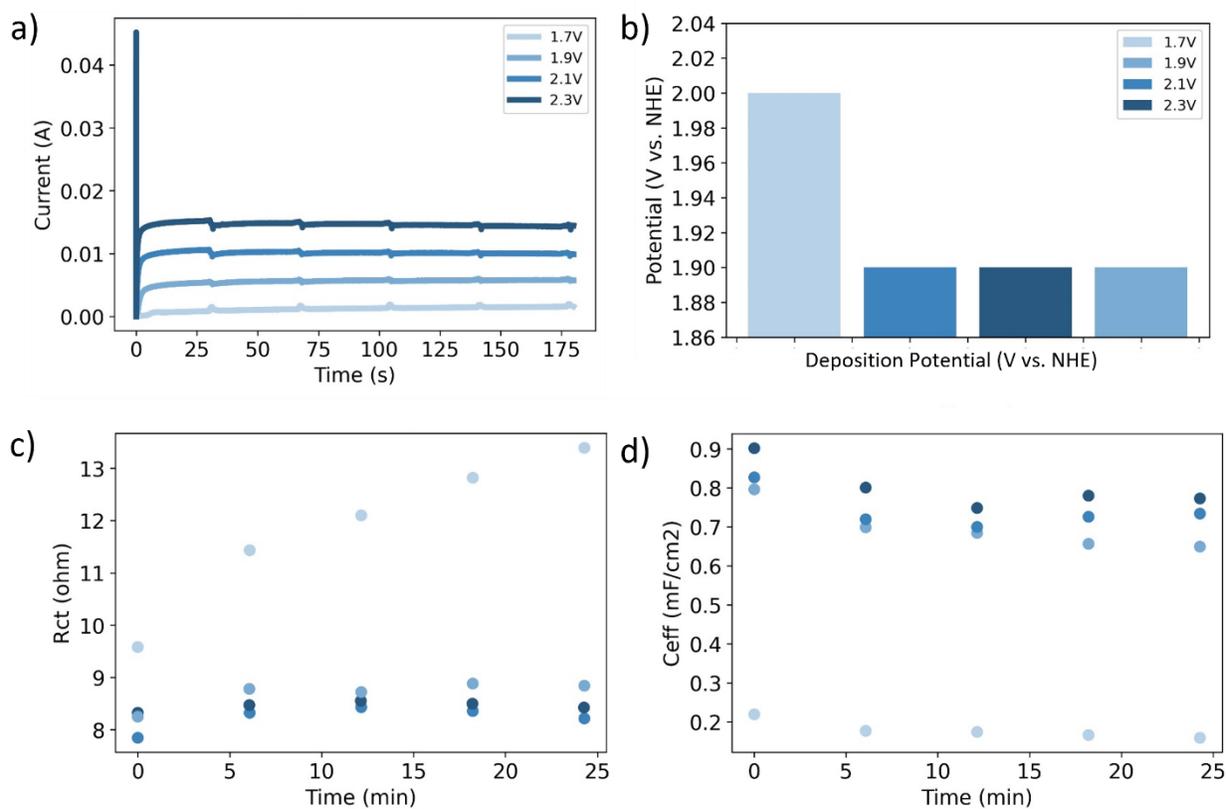

**Figure 3 -** *a) Current response of repeat Co and Pb-nitrate solution deposited samples at the indicated potentials, and b) corresponding OER potential measured at 5 mA/cm² in $H_sSO_4$ solution of pH = 2. Corresponding c) $R_{ct}$ and d) $C_{eff}$ values extracted from Randles circuit fit to the EIS spectra obtained at the time indicated during the OER.*

The total electrodeposition time becomes an important control parameter to ensure adequate loading of the catalyst material. Of concern during OER performance evaluation is the propensity of the mixed metal-oxide catalyst to undergo redeposition, should the metal ion concentration within the electrolyte be at a sufficient concentration.[39] This in turn interferes with the evaluation of the true OER overpotential metric. Ideally, the catalyst loading is large enough to evaluate proper OER kinetics, but also minimal in loading so that the concentration of metal ions in the solution remains low during material dissolution to minimize redeposition during the OER process. To decouple these two processes during OER, we utilized EIS with distributed relaxation time (DRT) analysis to elucidate the electrochemical processes.[48,49] We electrodeposited $CoPbO_x$ catalysts at 1.9 V vs. NHE for 3 minutes and 10 minutes, and performed OER for 30 minutes at 5 mA/cm$^2$ before performing EIS. The extracted DRT parameters (*Figure 4 a,b*) reveal that after 3 minutes, a single DRT peak centred at $\tau = 10^{-5.3}$ s dominates the spectrum, and a Nyquist model fit of a single Randles circuit *(R1-(P1/R2))* indicates the presence of primarily a single electrochemical process. This can be attributed to the OER reaction. A second much smaller shoulder exists at $\tau = 10^{-3}$ s, which is attributed to the redeposition process. For the 10-minute electrodeposited catalyst, both the Nyquist plot and the DRT indicate multiple electrochemical processes, which are suspected to be electrodeposition processes. The DRT displays a strong shoulder centred at $\tau = 10^{-1.8}$ s, indicating a larger contribution to the impedance of a second electrochemical process, and suggesting that a more appropriate Randles circuit of *(R1 – (P2/R2) – (P3/R3))* represent the electrochemical system. It is also of note that the primary OER peak centre shifts to $\tau = 10^{-4.7}$ s. The reason for this occurrence is not yet concluded but is likely related to either a change in the polarization resistance ($R_{ct}$) or the CPE element. The specific nature of this change will be investigated in future work.

To confirm these conclusions, two control experiments were performed to confirm the identity of the observed DRT peaks. The resulting DRT fit-EIS spectra for replicate 3-minute electrodeposited catalyst materials were subjected to a) OER in a different supporting electrolyte (0.1 M $HNO_3$) (*Figure S6a*), and b) OER characterization in $Co^{2+}$ / $Pb^{2+}$ nitrate spiked (0.1 M total metal ions) solution to mimic the scenario of a high dissolution of metal ions during the OER process (*Figure S6b*). OER performed in the 0.1 M $HNO_3$ displayed a single DRT peak at $\tau = 10^{-5.5}$ s which we attribute to the OER process. A second small peak is observed at $\tau = 10^{-1.7}$ s, which is at the approximate peak center position of the suspected re-deposition process. This is confirmed by the metal ion-spiked solution, where it is clear that the presence of metal ions dramatically shifts the DRT peaks to represent a host of new electrochemical processes at the lower $\tau$ regions. From these control tests, these lower frequency $\tau$ regions are determined to be the re-deposition electrochemical processes. Clear identification of these peaks and their specific corresponding electrochemical phenomena are outside the scope of this work. From this analysis, it is clear that a 3-minute deposition is ideal to minimize the metal-ion loading in solution and minimize re-deposition during electrochemical OER performance evaluation.

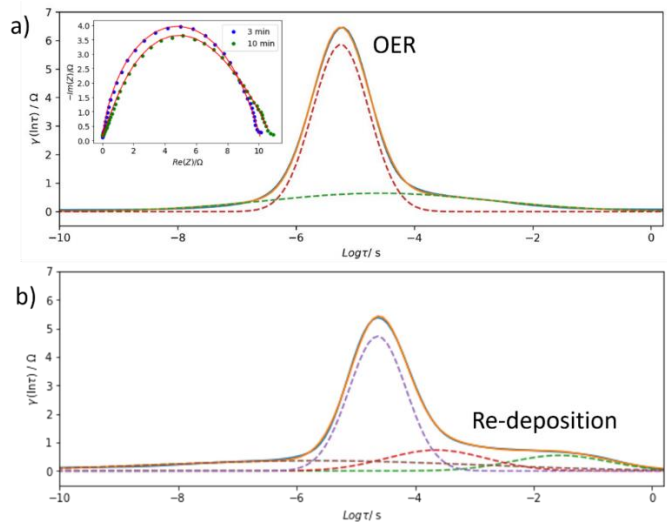

**Figure 4** - *DRT fit of the OER performance data after 30 minutes for samples prepared at a) 3-minute electrodeposition and b) 10-minute electrodeposition. Inset- Corresponding Nyquist plots of 1.9 V vs. NHE deposited $CoPbO_x$ OER catalysts after 30 min OER.*

*Electrochemical Performance Evaluation Protocol*

Of critical importance when attempting to optimize an experimental parameter via an active learning algorithm is the need to understand the experimental variability. Electrocatalysis OER measurements are prone to variability given the nature of the dynamic environment in which the experiment takes place.[50] Best practices address reactant transport limitations through the use of a rotating disk electrode to ensure controlled delivery of the reactants to the catalyst surface, and in turn, minimizes the residence of bubbles on the catalyst surface during the reaction. We do not have these capabilities in our self-driving laboratories due to the automation complexity of sample preparation. As shown in *Figure 2*, our electrochemical cell is static with no flowing electrolyte, which means bubble generation creates perturbations in the measured voltage. This can create variability in the current response. To determine the level of expected variability, multiple repeat $Co^{2+}$ / $Pb^{2+}$ samples were synthesized and evaluated with three different applied OER currents (1, 5, and 10 mA/cm$^2$) over 30 minutes of OER. ***Figure S7*** highlights these results. The sample-to-sample performance of these repeated experiments is relatively consistent, which is to be expected, albeit the measurement of the exact overpotential metric increases in variances as the current increases, as there is a more vigorous generation of bubbles. The overpotential mean and standard deviation of each current is 1 mA/cm$^2$: $\mu = 1.79$ V, $\sigma = 2.9$ mV; 5 mA/cm$^2$: $\mu = 1.87$ V, $\sigma = 4.3$ mV; and 10 mA/cm$^2$: $\mu = 2.09$ V, $\sigma = 14.9$ mV.

*Optimization Algorithm Development – Experimental Variability*

Based on the overpotential variability inherent in the OER performance measurements, we were determined to quantify how this variability impacts the performance of the machine-learning optimizer, in-order to influence our choice of optimizer for deployment in the self-driving laboratory. An extensive amount of optimizers have been developed and demonstrated successfully within self-driving laboratories.[26,27,51–53] Benchmarking studies have also been performed on various existing datasets, revealing how a specific optimizer performance is dependent on the data to be optimized, and a 'one size fits all' approach will not yield the best performance.[10,54] A majority of these optimizers operate in principle based on Bayesian optimization – utilizing experimental data as prior knowledge beliefs, and updating said beliefs with each successive experiment towards a specific objective function (maximization of a user-defined performance metric). However, the optimizer has a plethora of variables and hyperparameters that can be tuned and adjusted, each of which will impact how said optimizer performs within the experimental parameter landscape. Utilizing an existing electrocatalyst literature dataset from *Rohr & Stein et. al*[8], we explored the effectiveness of various optimizer models, hyperparameters, and acquisition functions when synthetic experimental variability. We chose this dataset based on its similarity to the type of experimental data generated from the previously discussed self-driving laboratory as part of this work. Our framework is based on the open-source optimization library BoTorch[53], a Gaussian Process (GP) probabilistic model optimizer built on Pytorch that offers a flexible and modular design, which allows users to easily customize the Bayesian optimization pipeline by plugging in different components such as acquisition functions, surrogate models, and optimizers.

Acquisition functions play a crucial role in optimizing algorithms and data queries in noisy observation environments. A detailed explanation of the optimization process and the role of the acquisition function has been covered in previous studies.[55,56] Acquisition functions determine the next query point by balancing exploration and exploitation. In this study, we investigated the effects of noisy data using upper-confidence bound (UCB) and expected improvement (EI) as our acquisition functions, and a combination of noisy-expected improvement (NEI) with various acquisition samplers (Quasi-Monte Carlo, Monte Carlo, and analytical) to enhance our analysis. These were implemented utilizing BoTorch model libraries of SingleTaskGP and FixedNoiseGP, which are single-output GP models. For hyperparameters of GP models, we choose Matérn52 kernel covariance function (with a high smoothness value, 2.5) and Automatic Relevance Determination (ARD)[57,58] to determine the relative importance of the inputs. A brief comparison of other Matérn kernels performance is shown in the *Supporting Information* (**Figure S8)**. EI balances the exploitation of solutions using the likelihood of improvement compared to the current best solution observed thus far, with the exploration of others which may be a lower likelihood of performance improvement, but provide the most information of the parameter space (ie. points with the highest variance). EI quantifies the overall expected improvement compared to the current best solution - $EI(x) = E[\max(f(x) - f^*, 0)]$, where $f(x)$ is the function to be optimized, $f^*$ is the current best-observed value of the function, and $E$ denotes the expectation with respect to the posterior of the function. UCB considers both the mean prediction and the uncertainty in the surrogate model. The most promising point (ie. experiment) to perform next is determined via an evaluation of the upper performance bound at each point in the parameter space $\alpha_{UCB} = \mu(x) - \beta\sigma(x)$. The degree of exploration vs. exploitation can be adjusted via a tradeoff parameter $\beta$. We optimize the $\beta$ parameter with HyperOpt module (infill BO model) during the optimization process. The BoTorch model and HyperOpt model are consecutively connected and are capable of being tuned based on the input data spaces. Another acquisition function - probability of improvement (PI) - was initially explored as part of this study, but a simulated performance comparison to UCB and EI, shown in **Figure S9**, revealed worse performance in comparison and hence it was not pursued as part of this benchmarking evaluation in noisy observation environments.

As mentioned previously, to evaluate the efficacy of these various models towards experimental output variability, we performed all optimizer evaluations on an OER catalyst data set from *Rohr & Stein et. al.* that was generated using high-throughput OER screening of mixed metal oxide samples produced via inkjet printing of a combination of precursors followed by controlled calcination.[8] This data set represents urinary to quaternary compositions of six possible metal-oxide combinations comprising of the cations [Mn–Fe–Co–Ni–La–Ce] at discrete composition abundances of [0:0.1:1] with a single output metric of electrochemical overpotential. This totals 2344 composition samples in this data set we chose to benchmark. This dataset provides a varied distribution in overpotential from 0.7 V to 0.37 V, and in their study, they have indicated that the relative performances of BO algorithms over different datasets were observed to be consistent. This same publication also introduces four benchmarking metrics to compare the active learning algorithm performance, a subset of which we have utilized as a part of this study.

To explore the effectiveness of the models in the design spaces with a limited number of samples (mimicking an SDL), we initiated the BO algorithms with one randomly selected sample and one query sample is picked from the pool in each iteration. The total pool of undiscovered data points can be expressed as $(\vec{x}\,;\,y) \in D$, whose input features $\vec{x}$ are all discrete data points and are made available for evaluation by the acquisition functions. Discrete material design spaces are chosen due to the resolution of the experimental set-up and overpotential measurements. The BO algorithms and the acquisition functions to be compared are evaluated in 20 campaigns of cycles with 20 independent random seeds. The aggregated performances of the BO algorithms were derived from 20 averaged runs with reporting mean and standard deviation of the results. We reduced the chances of landing on more informative samples by randomly selecting the first sample and running the algorithms in campaigns and comparing the mean and variance of the models. The Top% metric is a percentage of the top 30 catalysts that were selected from the entire parameter space.

$$Top\%(i) = \frac{Discovered\ top\ candidates\ [0,30]}{Number\ of\ top\ candidates\ to\ achieve\ (30)}$$

Enhancement Factor (EF) indicates how well the algorithm performs in finding the top candidates compared to the random baseline in each learning cycle ($i$) and is shown as:

$$EF(i) = \frac{Top\%_{Bo}(i)}{Top\%_{random}(i)}$$

The Acceleration Factor (AF) indicates how fast an algorithm can reach a defined number of top candidates. This parameter is an especially important metric for the comparison of expensive experiments, where it is ideal to accelerate the pace of exploration in as few experiments as possible.

$$AF(\#Top\ candidates = 30) = \frac{i_{BO}}{i_{random}}$$

The challenge associated with working with experimental data is they exhibit either homoscedastic or heteroscedastic behaviour. With this dataset, we added homoscedastic (NEI) or synthetic heteroscedastic experimental noise (SEN) to the overpotential outputs and evaluate the previously discussed algorithms, hyperparameters, and acquisition functions to gain an understanding of the strengths and limitations of such algorithms as they attempt to accommodate noisy observation environments. Homoscedastic noise is when the variance of the noise is constant across the entire dataset. The assumption of constant variance may lead to biased parameter estimates and emphasize the regions with low variance and underemphasize the regions with high variance.[60] The impact of heteroscedastic noise, a non-constant variability in the optimization metric across the dataset, has been reviewed to understand the challenges caused by experimental noise, which we present via SEN treatment of the output data. In this case, different levels of variability are present in different regions of the search space, which can lead to difficulties in accurately estimating the uncertainty and making informed decisions about the next query sample in a Bayesian optimization campaign. This is a better representation of true experimental noise as the variance of the noise changes concerning the input training data and real-world data which will apply more pressure for the BO models to perform. To evaluate the optimizer performance, we added 5, 10, and 15 mV variance to the overpotential, using the original data value for each as a mean value (NEI) or a random distribution of mean 0 (SEN), and trained the models based on this synthetic experimental variability.

The summarized results of this benchmarking study are shown in **Figure** 5. First, we performed a direct comparison of optimizer performance with no variability applied to the overpotential of the experimental dataset (ie. the measured dataset is treated as non-variable). This serves as a baseline comparison to evaluate EI and UCB acquisition functions (and associated β values of 0.05, 1.4, and 20). As mentioned previously, we utilized HyperOpt to determine an optimal value of β. Through HyperOpt, we identified β = 1.4 as the optimal β value in our study, on average able to identify the top 30 catalyst candidates with the lowest overpotential in 150 cycles. This is compared to UCB $_{(low\ β)0.05}$ (180 cycles), UCB $_{(high\ β)20}$ (225 cycles), and EI (220 cycles). It is important to note that the standard deviation of the aggregated model performance in each case is significant (UCB $_{(low\ β)0.05}$ = 96 cycles, UCB $_{(optimum\ β)\ 1.4}$ = 83 cycles, UCB $_{(high\ β)20}$ = 60 cycles, EI = 100 cycles) and is a strong representation of variability in these stochastic approaches. From the evaluation of the Top(%) benchmarking metric alone, it is difficult to determine a statistically significant superior optimizer. Evaluating the EF (**Figure 5b)** and AF **(Figure 5c)** provides significant insight into the applicability of these methods for our SDL. While EI performs worse compared to the UCB optimizers based on the previous metric, when looking at the efficiency of the optimizer it performs on par with that of UCB $_{(low\ β)0.05}$, in particular early on in the optimization cycle (ie. it initially learns at comparative rates). This is in contrast to UCB $_{(high\ β)20}$, which takes significantly longer to reach comparative learning rates as evident from the sluggish EF and AF over the first 100 learning cycles. This is expected behaviour, as UCB with higher β values compared to optimum value are expected to exhibit greater exploration behaviour and requires a greater number of learning cycles to develop efficient models compared to lower β values, which are more exploitative. As the data is more spread around the global minimum, exploitation is more important to the performance of the models and as a result lower β values have better performance at the early stages of learning, which diminishes compared to optimal β values as the number of learning cycles increases. As expected, all optimizers outperformed a completely random selection which we used as a baseline.

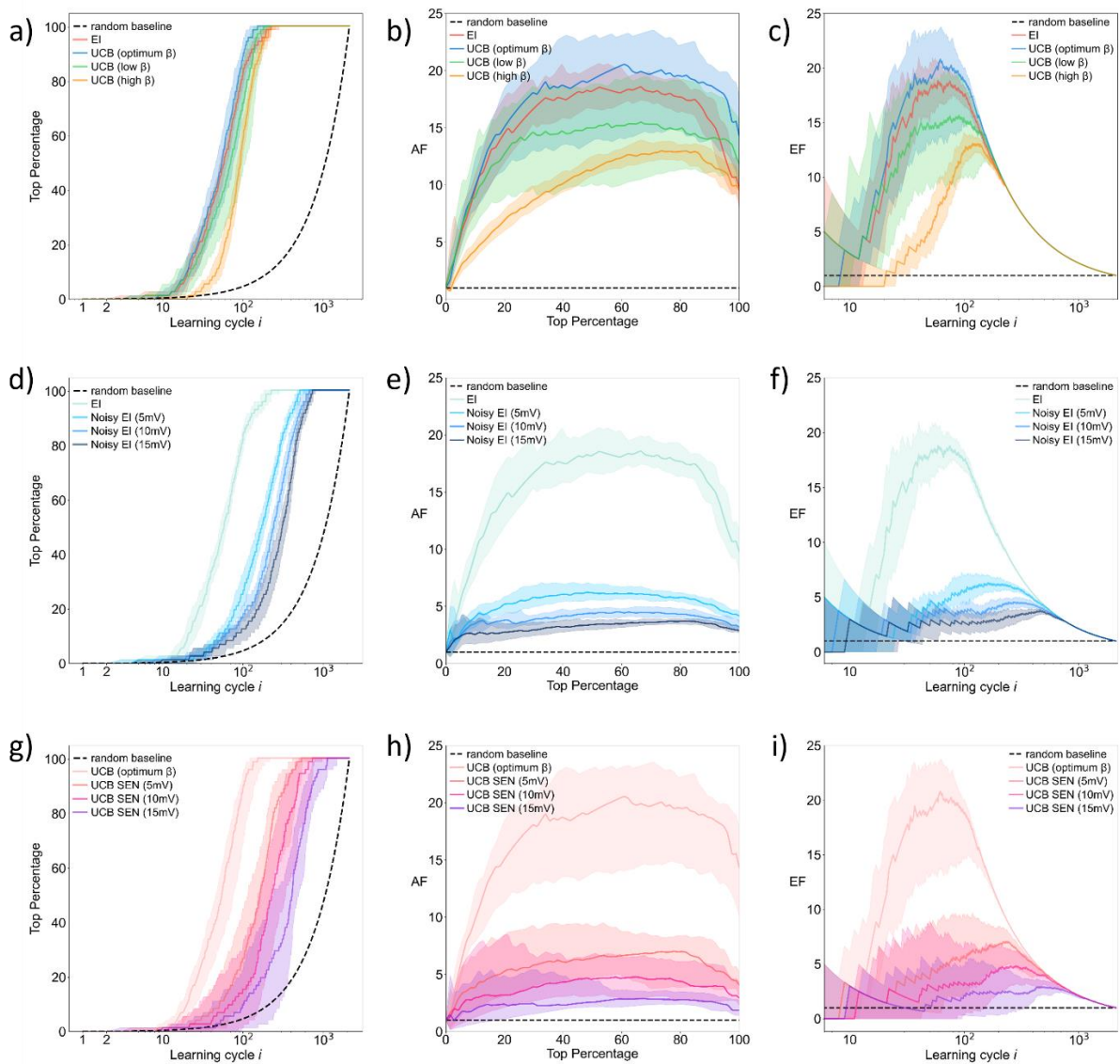

**Figure 5** – *Benchmark performance comparisons for Top%, AF (acceleration factor), and EF (enhancement factor) of EI and UCB (various β) acquisition functions, no experimental noise (a-c), 5, 10, 15 mV homoscedastic noise applied to EI (labelled Noisy EI) (d-f), and heteroscedastic noise applied to UCB (labelled SEN) (g-l), respectively. This benchmarking study was applied to experimental data from [8], as detailed in the body.*

Under this same framework, we evaluated the addition of variance to the overpotential, with the results shown in **Figure 5d-f** (NEI) and **Figure 5g-i** (SEN), plotted alongside UCB $_{(optimum\ β)}$ = 1.4 and EI with no noise, for comparison. As expected, as the variability of the measurement output increases from 5, 10, to 15 mV, the optimizer performance subsequently decreases by greater margins. In comparing NEI and Top% (using 100% as the benchmark metric), we see an increase in the learning cycles from a mean of 224 learning cycles with no artificial noise to 512, 670, and 734 learning cycles, respectively. This also comes with a greater degree of standard deviation, which is summarized in **Figure 6.** A similar comparison using UCB $_{(optimum\ \beta)1.4}$ with 5, 10, and 15 mV variability (SEN) applied to the overpotential results in a similarly expected decrease in performance compared to the non-experimental noise case. In particular, UCB performs the best when no artificial noise is added (mean 152 cycles), but fairs worse compared to NEI when artificial noise is added (537, 738, and 1129 learning cycles, respectively). Furthermore, the standard deviation of SEN is significantly greater compared to the variability-added evaluations with NEI, which is expected given the heteroscedastic nature of this simulated noisy experimental environment.

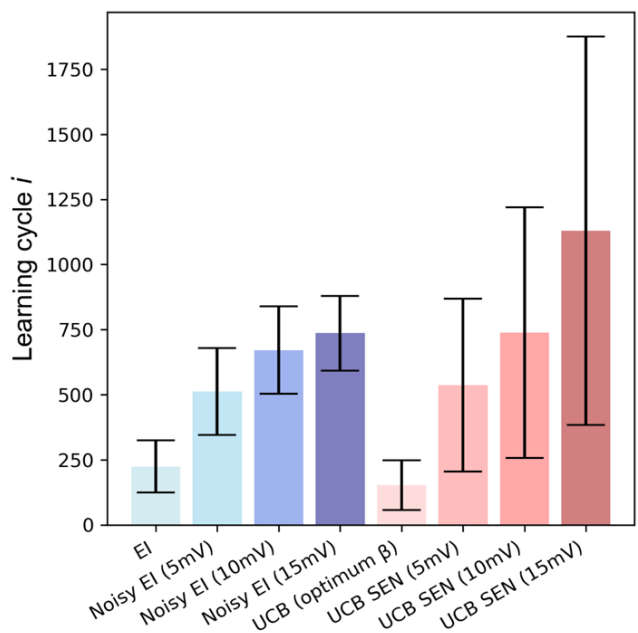

**Figure 6** – *A summary of all benchmarked Top(%) tests, comparing the mean number of learning cycles required to identify 100% of the top 30 catalysts out of a possible 2344. Error bars define the standard deviation determined from the aggregation of 20 simulated campaigns.*

In conclusion, this benchmarking evaluation reveals that to have any confidence in our global optimization given the performance variability during our catalyst evaluation, we must limit our variability to 5 mV. Based on this, we have chosen to use a performance evaluation current of 5 mA/cm$^2$, which previously was determined to give σ = 4.3 mV, as an acceptable metric between experimental variability and relevant catalyst performance evaluation. It is clear, however, the great importance to limit sources of experimental variability as they have a drastic impact on the overall optimizer efficiency. If the goal is to reduce the number of learning cycles (ie. expensive experiments), even small amounts of experimental variability can generate large increases in the required learning cycles. While our small study evaluated experimental uncertainty on the output variable (the variable attempting to be optimized), other studies have been reported with a robust methodology and exploration of input experimental variables[51], which was not considered in this benchmarking study.

*Optimization – Deployment on self-driving laboratory*

We performed two separate optimization campaigns using the self-driving laboratory outlined in this work, with the design considerations mentioned in prior sections. For each campaign, the electrodeposition target elements were Co, Fe, Mn, and Pb, using discrete concentrations of between 0 M and 1 M, with an increment step size of 0.1 M, generating an experimental landscape of 256 possible different deposition solution compositions. The measured overpotential after 30 minutes of applied constant current (5 mA/cm$^2$) for each catalyst run is shown in **Figure 7a**, with each run plotted as a function of the electrodeposition composition (solution molarity of Co$^{2+}$, Pb$^{2+}$, and combined (Mn$^{2+}$ + Fe$^{3+}$)) in **Figure 7b,** and associated images of each catalyst in **Figure S10.** In summary, over both campaigns, our self-driving laboratory can identify successful regions for catalyst synthesis within 5 experiments, and identification of the best-performing region (ie. lowest overpotential) in less than 10 experiments, which corresponds to a solution composition range of Co$^{2+}$: 0.4 – 0.6 M, Pb$^{2+}$: 0.4 – 0.6 M, Mn$^{2+}$/Fe$^{3+}$: 0 – 0.1 M. From a scientific standpoint, the experimental results obtained with this self-driving laboratory are corroborated with those results in scientific literature. In particular, previous studies have demonstrated that a 1:1 combination of Co$^{2+}$:Pb$^{2+}$ with the addition of a small amount of Mn$^{2+}$ [35] or trace amounts of Fe$^{3+}$ [37] in the deposition solution leads to the best-performing catalysts within their study. Our study serves to demonstrate that with a self-driving laboratory and AI-guided optimization, the same results can be identified on the order of hours.

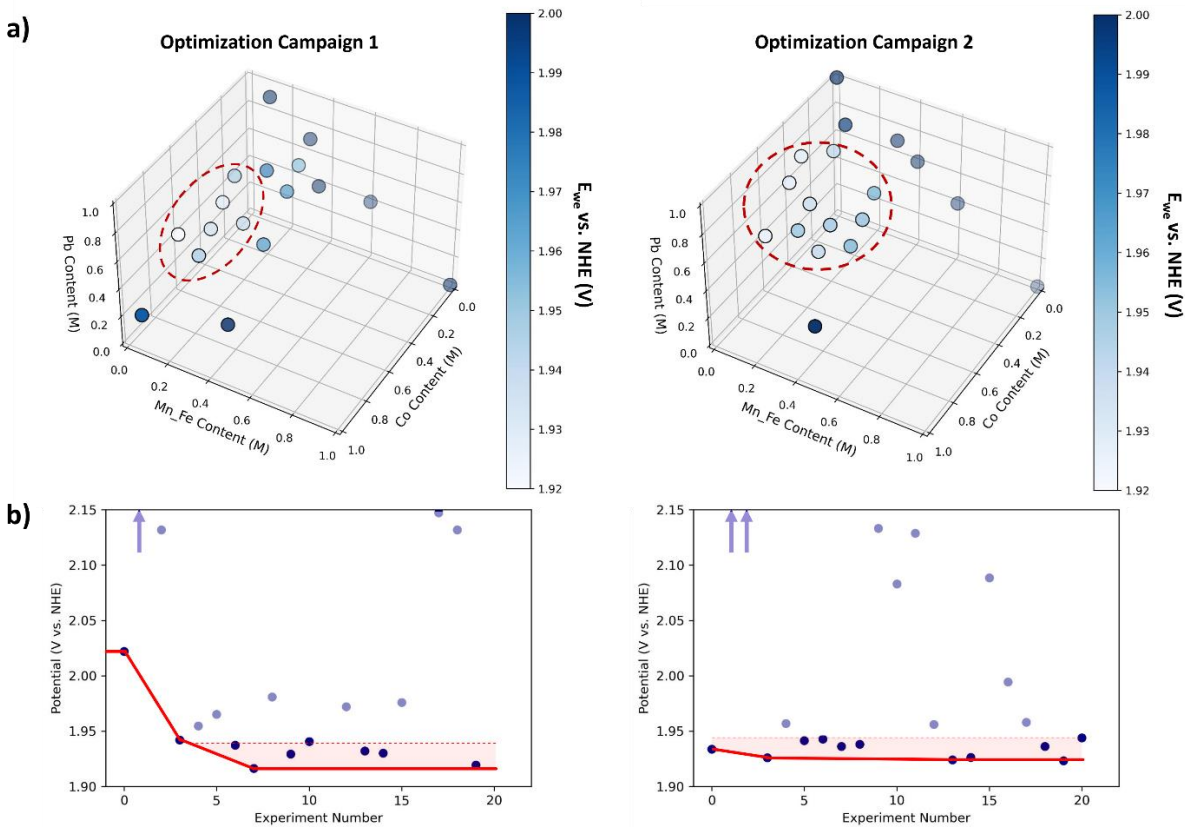

**Figure 7 -** *a) 3D plot of Co, Pb, and Fe + Mn contents as a function of OER potential for all runs performed during the 20-run optimization. The best-performing regions for each optimization are circled in red. b) Potential vs. experiment number for the optimization process. The red line represents the current optimum catalyst during the campaign, and the shaded region corresponds to the circled region in a).*

While not a focus of the optimization, stability was also a critical metric that we measured during the OER reactions. Stability optimization coupled with overpotential optimization will be the focus of future work. Of the data gathered during these optimization campaigns, the EIS data and associated equivalent circuit model component values reveal critical information about the relationship between the electrochemical deposition solution composition and the catalyst stability. All EIS data was analyzed from both experimental campaigns discussed above, and the percentage change of $R_{ct}$ and $C_{eff}$ as a function of each element concentration was analyzed for material-performance correlations. The Nyquist plots obtained at t = 3 min and t = 30 min are shown in **Figure S11**, and show the change in the Nyquist plots throughout a 30-minute OER experiment. As a majority of the top-performing catalysts were comprised of Co and Pb, **Figure 8a-d** plot the percentage change from t = 3 min and t = 30 min for $R_{ct}$ and $C_{eff}$ versus overpotential with a hue representing the concentration of a specific element. Evaluation of $C_{eff}$ change (**Figure 8a,c**) reveals that a heavy Co composition (> 0.6 V vs. NHE) does minimize the change in $C_{eff}$, and in term is an indicator of greater stability compared to those catalysts with lower Co content. Pb concentration, on the other hand, appears to have weak or no correlation given the even distribution of the data points. Changes in $R_{ct}$, however, appear to not correlate with elemental composition for both Co and Pb, outside of low concentrations (< 0.3 M) of either element leads to drastic changes in $R_{ct}$, as would be expected given the higher rates of dissolution and/or poor intrinsic OER activity when catalysts low in Co or Pb are synthesized and tested. These results and correlations are highly promising, albeit more work is necessary to further elucidate the scientific understanding behind these results with more targeting experiments to further probe the claims made here. Regardless, these early results demonstrate how *proxy* experimentation can be developed and utilized in the place of more exhaustive and expensive characterization techniques for purposes of optimization and understanding.

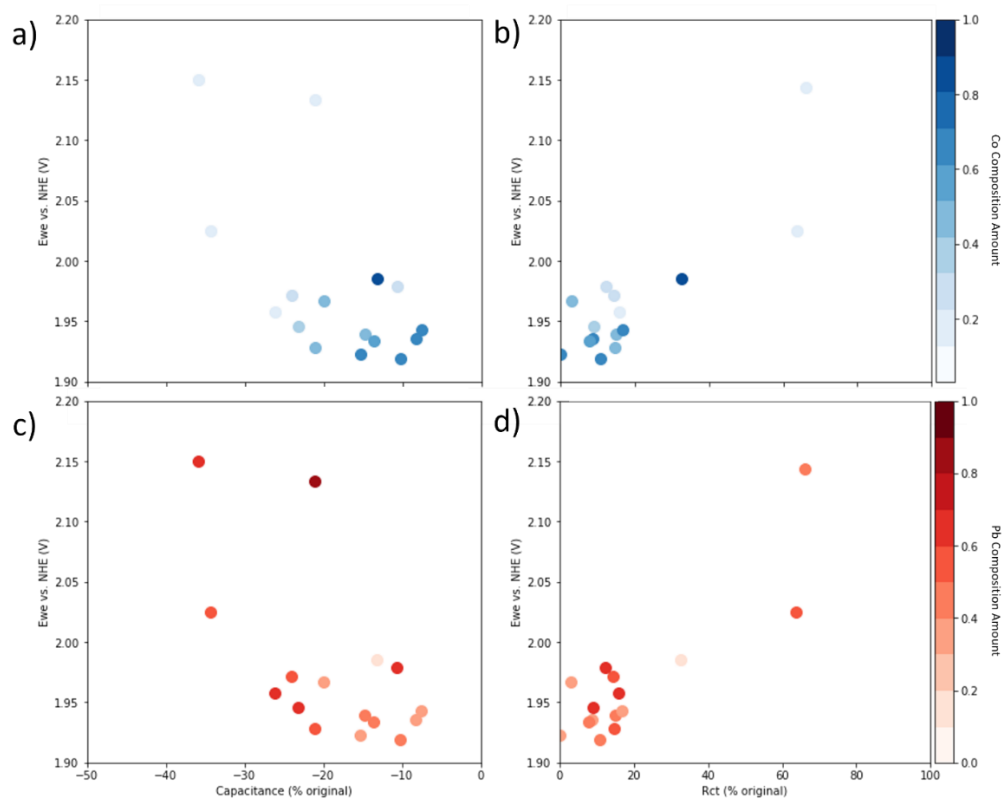

**Figure 8** - *Percentage change in $C_{eff}$ and $R_{ct}$, as determined from the Randles ECM for EIS at t = 3 min and t = 30 min, as a function of Co metal content (a, b) and Pb metal content (c, d) in the CoFeMnPbO$_x$ catalyst material. Associated Nyquist plots are shown in Figure S11.*

*Conclusions*

In conclusion, this study has demonstrated the effectiveness of a self-driving laboratory in optimizing mixed-metal oxide electrocatalysts for acid-stable OER. The incorporation of an experimental optimizer enabled us to quickly identify an optimal electrodeposition procedure with equimolar Pb to Co ratio, and small dopings of either Fe or Mn for improved overpotential performance, with this finding corroborated by the literature. Additionally, we explored the use of proxy experimentation as a valuable tool for designing experimental workflows when expensive characterization equipment is not available, and have shown an initial view of generated correlations between elemental composition and pseudo-acid stability through electrochemical measurements alone. While the concepts of proxy experimentation presented in this work by no means can replace rigorous material characterization methods, they offer an opportunity to utilize versatility and carefully designed experimental protocols in unconventional ways to gather some scientific intuition and/or information that would otherwise not be possible. This is a concept we plan to further develop and report on in future work. We emphasize the importance of reproducibility and validity in measurements and have presented our approach to improving this methodology as it relates to this work. Moving forward, we plan to expand our capabilities with onboard characterization to incorporate other metrics and improve our scientific understanding of property-performance relationships. Furthermore, we aim to explore other electrodeposition parameters as they relate to other electrochemical systems. We encourage other researchers to critically consider the use of self-driving laboratories and proxy experimentation to optimize their experiments and elucidate performance information.

**Supporting Information**

*Experimental*

The electrodeposition and electrochemical performance evaluation are discussed within the body of this work. Anodic electrodeposition conditions employed previous literature procedures, and utilized hydrate forms of nitrate salts in total concentrations between 0 – 1 M in 0.1 M nitric acid ($HNO_3$, 70% ACS reagent diluted with 18 MΩ Millipore $H_2O$). Electrodeposited samples were of the oxy-hydroxide form ($CoFePbMnO_x$) and were synthesized using cobalt(II) nitrate hexahydrate ($Co(NO_3)_2 \cdot 6H_2O$, 99.999% trace metals basis, Sigma Aldrich), lead(II) nitrate ($Pb(NO_3)_2$, 99.999% trace metals basis, Sigma Aldrich), manganese(II) chloride tetrahydrate ($MnCl_2 \cdot 4H_2O$, 99.99% trace metals basis, Sigma Aldrich), and iron(III) nitrate nonahydrate ($Fe(NO_3)_3 \cdot 9H_2O$, 99.999% trace metals basis, Sigma Aldrich) from a 1 M stock solution, added and diluted to 10 mL of the optimizer-specified concentration. An automated dispensing unit with an accuracy of 0.05 mL was used to prepare the electrodeposition samples, with a mass balance used to confirm the final solution concentration to +/- 1% targetted molarity. All electrodeposition procedures were performed in a 3-electrode cell with pre-cleaned fluorinated tin oxide (FTO) glass slide used as the working electrode and deposition substrate (MSE supplies, 2.2 mm thick, 7-8 Ohm/Sq TEC 7, 25 x 25 mm), an Ag/AgCl reference electrode (3M KCl, 6 mm, BaSi), and a cleaned graphite rod as the counter electrode. The electrochemical OER performance evaluation utilized a similar electrochemical set-up in a duplicate electrochemical cell to avoid cross-contamination, except used a Pt coil as the counter electrode as opposed to a graphite rod. The OER performance electrolyte was 0.01 M $H_2SO_4$ (99.999%, Sigma Aldrich). After each experiment, all cell contents were washed with 18 MΩ millipore water 10 times to remove trace metal concentrations (*See Figure S1*).

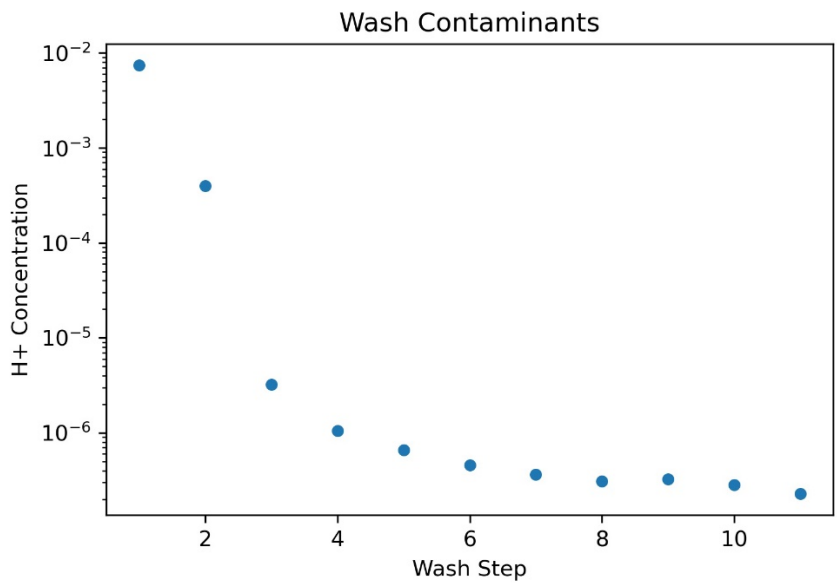

**Figure S1 -** *pH of deionized water after each successive wash step. At 8 successive wash steps ensure full cleaning of the cell and minimize contaminants to less than 1 ppm.*

*Proxy evaluation of catalyst stability*

The true evaluation of catalyst stability requires operating the catalyst in a representative environment until the catalyst is either fully deactivated or dissolved. We wanted to investigate if we could reduce the time required to obtain a relevant descriptor of catalyst stability and utilize this approach to obtain a *proxy* method for comparing catalysts within the same campaign as a means to reduce the experimental time required for the performance evaluation.

*Correlation of potential response as a proxy for catalyst stability*

**Figure S2a** shows the electrochemical voltage responses of the various catalysts tested. For each electrochemical curve, the voltage slope was extracted at times of 10, 20, and the full lifetime of the catalyst material. The full lifetime of the catalyst was taken at the inflexion point to higher voltages, which was visually confirmed as this inflexion point corresponded to when the thin catalyst layer was completely removed from the FTO substrate during OER. **Figure S2b** shows the extracted and truncated slopes as a function of time, and **Figure S2c** shows the correlation of the voltage slope vs. true catalyst lifetime, which we concluded to be a weak correlation of catalyst lifetime, with both 10-minute and 20-minute predictions having a mean average error (MEA) of > 20 minutes, or significantly worse for those catalyst materials exhibiting lifetime outside the bounds of the experimental data set.

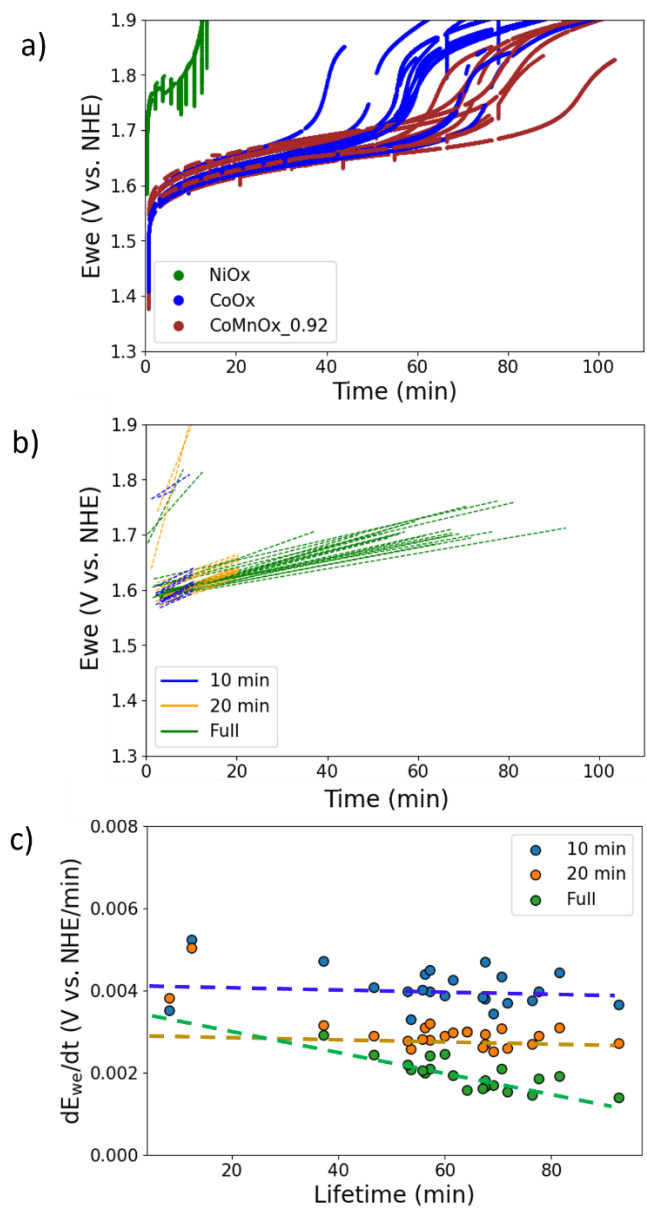

**Figure S2** - *a) Raw potential profiles of the indicated catalysts as a response to an applied current of 0.5 mA/cm$^2$, b) corresponding linear fits to the slopes within the as-indicated time range, and c) plot of voltage slope vs. the lifetime of the catalyst for each time indicator.*

*Correlation of $C_{eff}$ as a proxy for catalyst stability*

Subsequently, current hold EIS measurements were taken every 3 minutes during the entirety of the constant-current OER measurement until catalyst dissolution. A single Randles circuit with a constant phase element (CPE) was fit to each EIS spectra to extract circuit element values for $R_s$, $R_{ct}$, $C_{CPE}$ and α. Given the nature of these experiments (low current and thin-film catalyst materials), a single Randle circuit was found adequate to represent the obtained data. The interfacial capacitance of mixed-metal oxide surfaces, in particular those under dynamic OER conditions is difficult to truly evaluate and equate to the true electrochemical active surface area, as the true specific capacitance of our materials is impossible to accurately determine given the dynamic nature of the amorphous mixed-metal oxide material during OER. Hence we focus on establishing a capacitance term that is dependent on the contributions from the solution and polarization resistances.[62] The following equation was used to extract an effective capacitance value for these relations, which we have termed $C_{eff}$ (effective capacitance) as a representation of the capacitive double-layer.

$$C_{eff} = \left( C_{CPE} \left( \frac{1}{R_s} + \frac{1}{R_{ct}} \right)^{\alpha-1} \right)^{1/\alpha}$$

$C_{eff}$ were plotted versus time, and we observed a near-linear fit to the corresponding data (**Figure S3a-c**). Linear fits were performed for each catalyst after 10 minutes, 20 minutes, and full lifetime measurements were performed to extract values of slope ($C_{eff}$/time). The slopes were then used as the dependent variable to fit a simple linear regression model as a predictor of catalyst lifetime, with the results shown in **Figure S3d-f**. Based on these results, we determine that the intermittent EIS during constant current and extraction of $C_{eff}$ change over time, and determination of the $C_{eff}$ slope throughout a 20-minute constant current OER measurement is an adequate *proxy* metric of catalyst stability – in particular a much stronger indicator compared to that of voltage change over time.

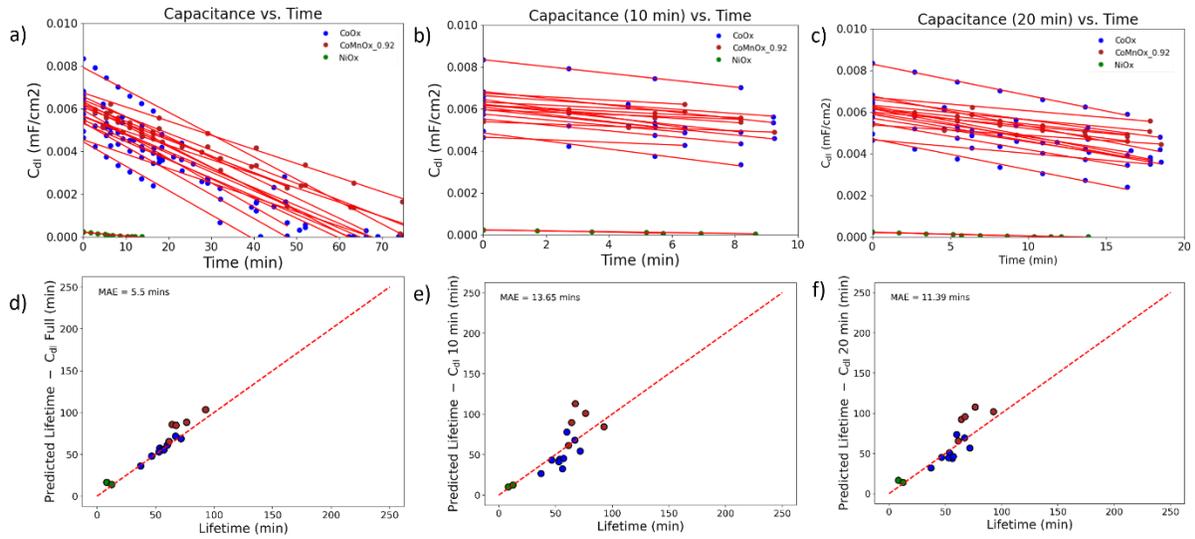

**Figure S3** – Effective *capacitance values fitted from EIS data vs. time during constant current discharge plotted and linearly fitted for a) full lifetime of the electrode, b) 10-minute lifetime and c) 20-minute lifetime. The linear fits and MAE correlations of real vs. predicted lifetime based on the linear regression fit are shown in d)-f) for the same timeframe.*

**Bubble Templating During Electrodeposition**

During anodic electrodeposition, bubbles are generated at the electrode surface which blocks the substrate sites from further deposition. These bubbles can occlude the catalyst surface, creating a templated, visibly non-uniform coated electrode surface. We standardized the electrodeposition procedure to include intermittent pulsing during the electrodeposition step to occur every minute to remove the bubbles from the surface. The result of this pulsed templating method is a visually smooth catalyst surface that is void of bubble templating, as shown in *Figure S4*. These intermittent pulses can be seen in the current response of the electrodeposition (*Figure S4b*), to which the system quickly reaches back to equilibrium after only a few seconds. The reproducibility of this method to produce similar catalyst thin films was demonstrated over 5 samples of the same deposition conditions, shown via the current response during deposition (*Figure S4b*) and associated EIS spectra obtained after 5 minutes of OER at 5 mA/cm$^2$ current.

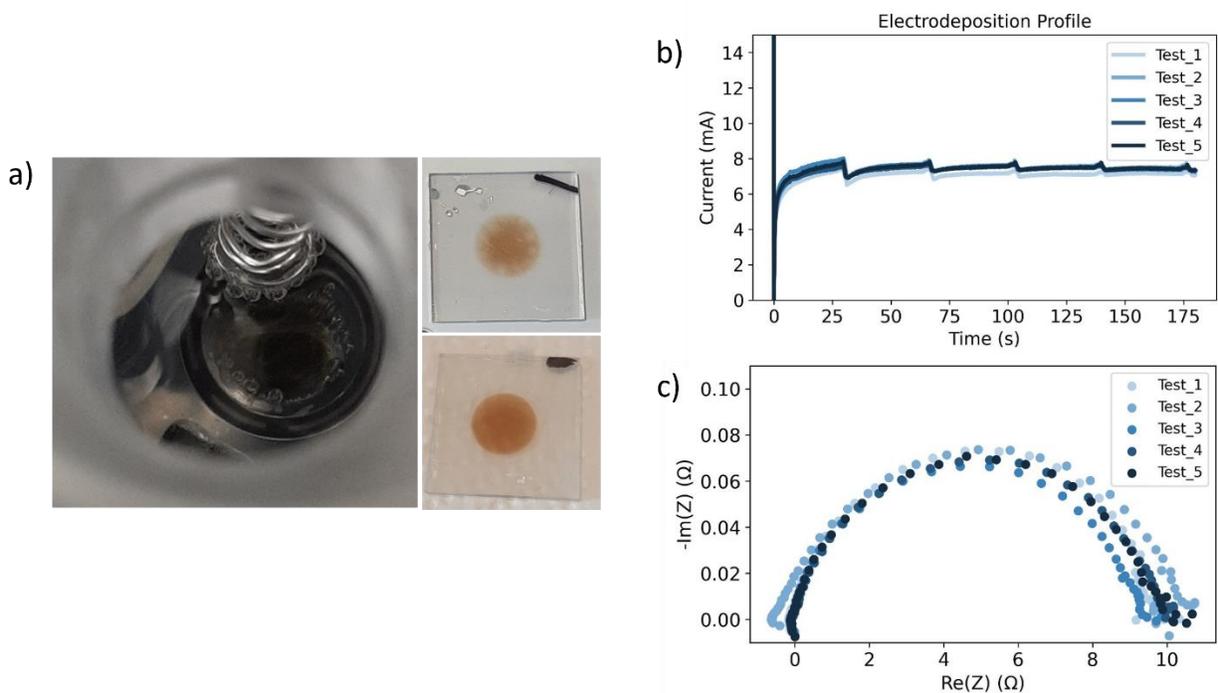

**Figure S4** - *a) Images of the bubble formation during electrodeposition of the catalyst material, with the inset showing the image of bubble templating (top) vs. a clean sample that was pulsed with liquid to remove bubbles during deposition to remove the bubbles (bottom). b) corresponding deposition profile and c) EIS after 5 minutes of OER, with current held at 5 mA/cm$^2$.*

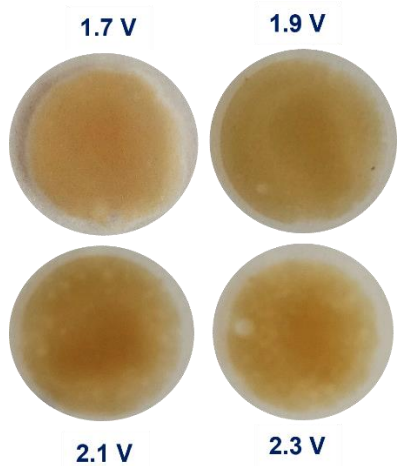

**Figure S5** - *Images of the electrodeposited samples from Figure 3.*

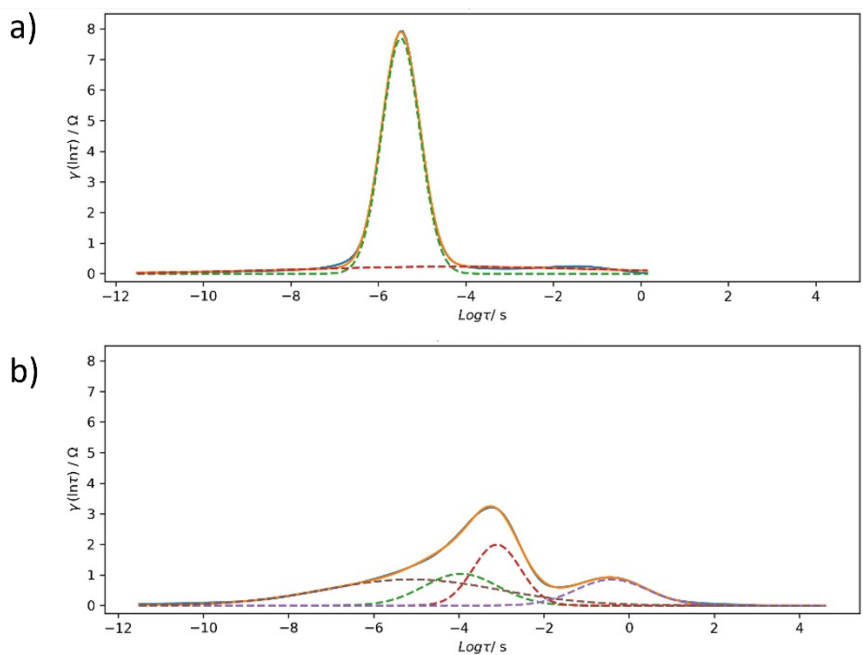

**Figure S6** - *a) DRT fits to the data for 3-minute electrodeposition and OER performed in a different supporting electrolyte (0.1 M HNO$_3$ as opposed to H$_2$SO4) and b) OER performed in the presence of Co$^{2+}$ and Pb$^{2+}$ nitrate, 0.1 M total metal ions.*

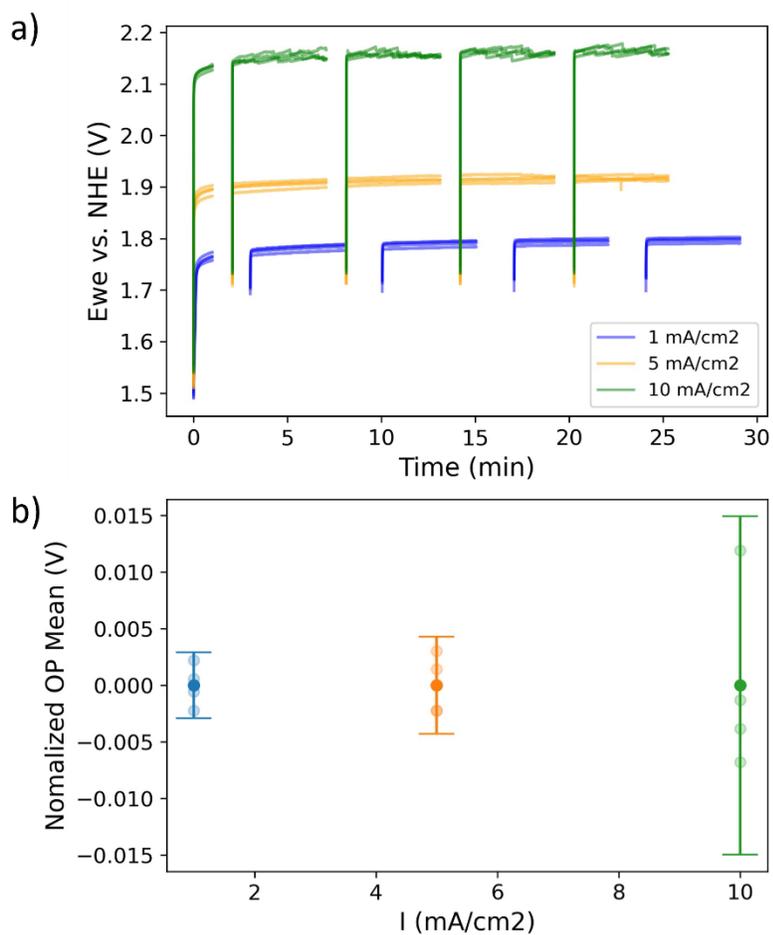

**Figure S7** – *a) CoPbO$_x$ catalyst samples deposited under the sample electrodeposition conditions, with OER measured at 1 mA/cm$^2$, 5 mA/cm$^2$, and 10 mA/cm$^2$. b) Corresponding overpotential (OP) mean and error bars to indicate the variability associated with each determining OER overpotential at each applied current. σ = 2.9 mV, 4.3 mV, and 14.9 mV, respectively.*

*Benchmarking Algorithms, Hyperparameters, and Acquisition Functions*

*Computing and Efficient Hyperparameter Tuning*

In this work, we have utilized open-source Bayesian optimization algorithms, BoTorch from Pytorch and HyperOpt. The simulated materials optimization was conducted in 20 campaigns on CUDA and CPU on a computer with 32 GB RAM and core i7-10850H CPU and 16 GB GPU. Running the Pytorch algorithms (BoTorch) on CUDA have the benefit of reducing the computer's high configuration requirements. In HyperOpt, the inner loop for optimization of hyperparameters, the number of optimization attempts is set to 100 and the exploration space is set to be from 0.1 to 100. This technique has enabled us in finding the best hyperparameters for parameters for the Bayesian optimization algorithm.

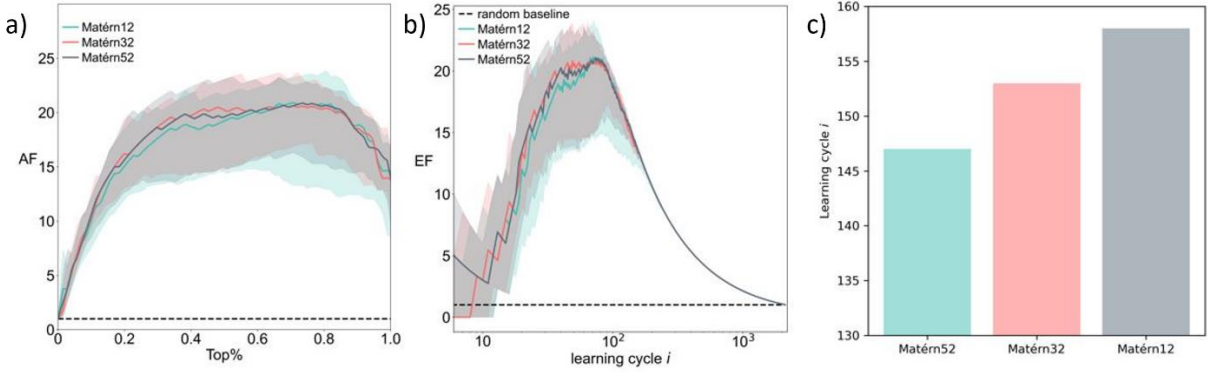

**Figure S8 -** *Benchmarking Algorithms, Hyperparameters, and Acquisition Functions. Comparing Matérn12, Matérn32, and Matérn52 kernels with a base smoothness value of 2.5, there is no statistical difference in performance for a) acceleration factor (AF) and b) enhancement factor (EF). A more complex pattern in data can be captured by kernel covariance function with higher smoothness values and as it is shown in c) lower learning cycles were required in Matern52 to reach and learn the top 30 candidates with lowest overpotentials in the whole dataset.*

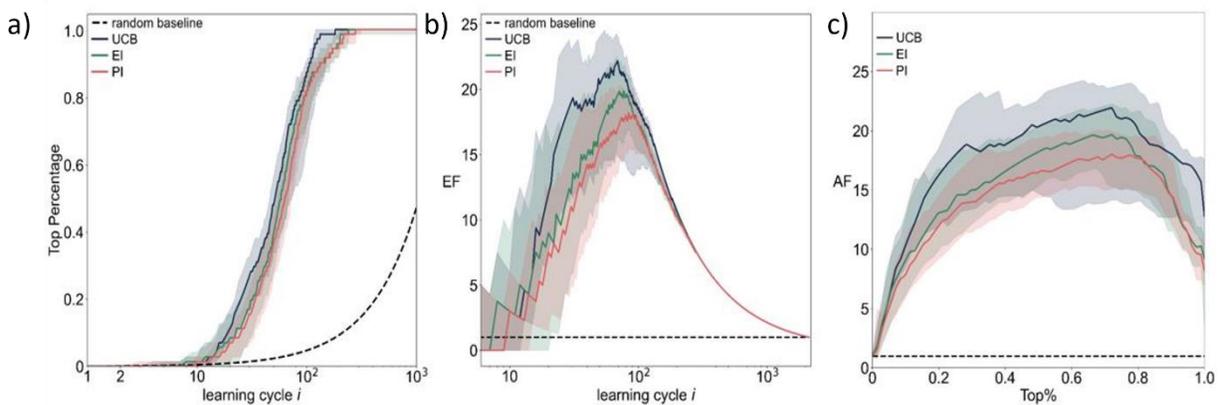

**Figure S9** – *Comparison of Acquisition Functions Without Existence of Noise. Comparison of the acquisition functions upper confidence bound (UCB, β = 1.4), expected improvement (EI) and probability of improvement (PI), no simulated experimental noise. This comparison is performed within the 2344 experimental dataset used throughout the text. By all metrics of a) learning cycle vs. top percentage (top 30 catalysts), b) enhancement factor (EF) and c) acceleration factor (AF), UCB outperforms both EI and PI.*

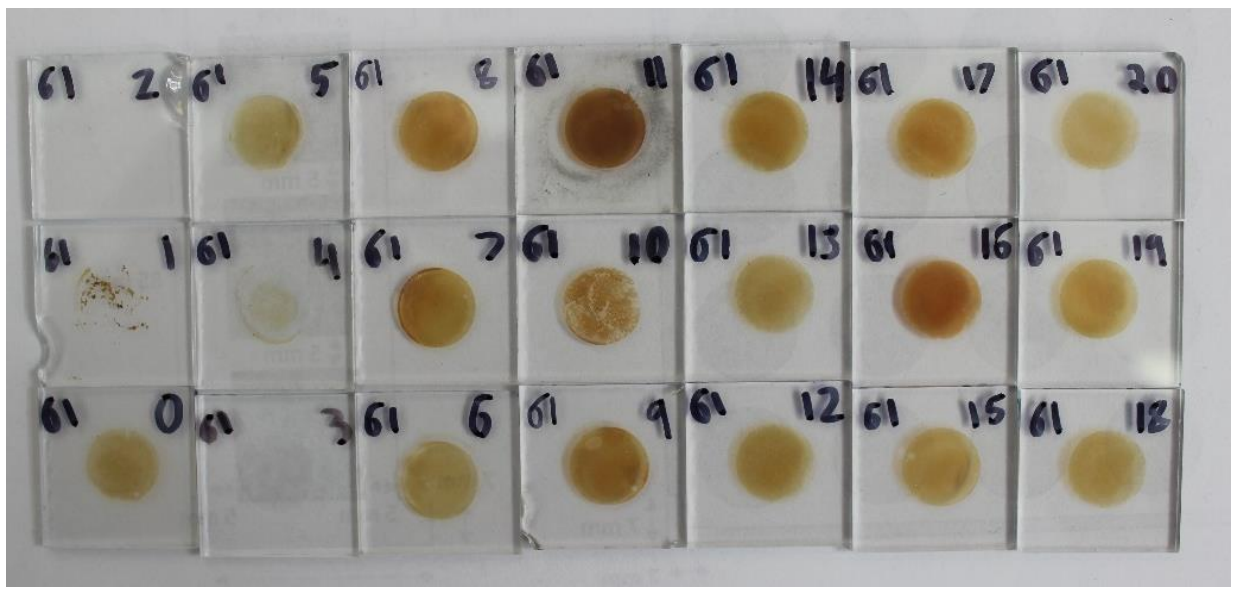

**Figure S10** – *Photographs of each catalyst sample after 30-minute OER at 5 mA/cm$^2$, produced from Optimization Campaign 1, with the sample (experiment number) on the top right of each sample.*

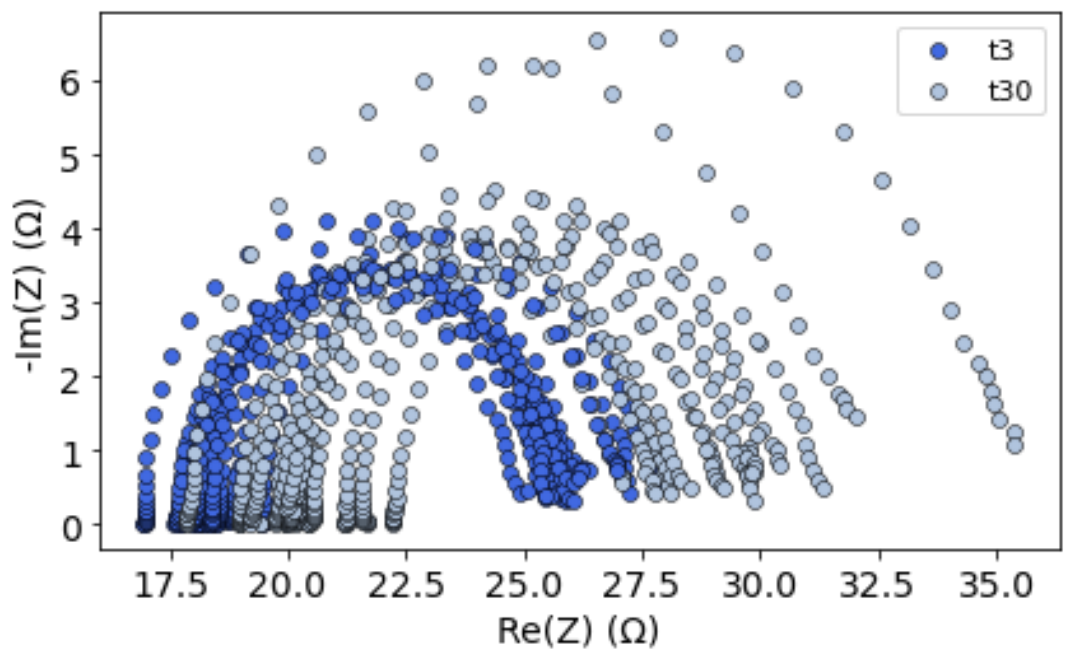

**Figure S11** - *Nyquist plots for t = 3 min and t = 30 min throughout a single optimization campaign.*